\documentclass{aastex}
\usepackage{emulateapj5}
\usepackage{apjfonts}

\def\calc{{\cal C}}

\def\cc{{\cal C}}

\def\crr{{\cal R}}

\def\llouv{\log L_{\rm OUV}}
\def\llx{\log L_{\rm X}}

\def\nudip{\nu_{\rm dip}}

\def\tt{\times 10}

\def\llnir{\log L_{\rm 1-25 \mu }} 
\def\llmir{\log L_{\rm 1-60 \mu }}
\def\llfir{\log L_{\rm 1-100 \mu }}
\def\llouv{\log L_{\rm 0.1-1 \mu }}
\def\llx{\log L_{\rm X}}

\def\lvdip{\log \nu_{\rm dip}}
\def\dd{\partial}

\journalinfo{The Astrophysical Journal Letters:}

\begin{document}

\title{Evolutionary Consequences of Dusty Tori in Active Galactic Nuclei}

\author{Jian-Min Wang\altaffilmark{1}, En-Peng Zhang\altaffilmark{1,2} and Bin Luo\altaffilmark{1,2}}

\altaffiltext{1}{Laboratory for High Energy Astrophysics,
       Institute of High Energy Physics, Chinese Academy of Sciences,
       Beijing 100039, P. R. China, wangjm@mail.ihep.ac.cn}
\altaffiltext{2}{Graduate School of Chinese Academy of Sciences, Beijing 100039, P. R. China}

\slugcomment{Received 2004 September 2; accepted 2005, May 18}
\shorttitle{Evolutionary Consequences of Dusty Tori}
\shortauthors{WANG, ZHANG \& LUO}

\begin{abstract}
Deep surveys of {\em Chandra} and {\em HST} (Hubble Space Telescope) show that active galactic 
nucleus (AGN) populations are changing with hard X-ray luminosities. This arises an interesting 
question whether the dusty torus is evolving with the central engines. We assemble a sample 
of 50 radio-quiet PG quasars to tackle this problem. The covering factors of the dusty 
tori can be estimated from the multiwavelength continuum. We find they are strongly 
correlated with the hard X-ray luminosity. Interestingly this correlation agrees with the fraction 
of type II AGNs discovered by {\em Chandra} and {\em HST}, implying strong evidence 
for that the AGN population changing results from the evolution of the tori. We also find that 
the frequencies of the dips around 1$\mu$m in the continuum correlate with the covering factors 
in the present sample, indicating the dip frequencies are adjusted by the covering factors. In 
the scenario of fueling black hole from the torus, the covering factor is a good and the dip 
frequency is a potential indicator of the torus evolution. 
\end{abstract}


\section{Introduction}
The polarized spectrum of NGC 1068 shows a broad H$\beta$ feature indicating the presence of a 
hidden active nucleus by a geometrically and optically thick torus (Antonucci \& Miller 1985). 
The opening angle is estimated about $74^{ \circ}\sim 96^{\circ}$ for Seyfert sample from the 
relative number of type II AGNs (e.g. Osterbrock \& Shaw 1988; 
Tovmassian 2001), similar to PG quasars from continuum (Cao 2005).
The unification scheme is successful in explanation of the Seyfert 1/2 dichotomy (Antonucci 1993).
However the strong unification scheme, namely a torus with a constant $H/R$, seems difficult in 
explanation of the multi-wavelength continuum (e.g. Wilkes et al. 1999). Deep surveys of 
{\em Chandra} and {\em HST} show the fraction ($P_{\rm II}$) of type II AGNs decreases with 
2-10keV luminosities ($L_X$) (Ueda et al. 2003; Steffen et al. 2003; Hasinger 2004, 
hereafter H04; but see a different explanation in Zhang 2005). 
An selection effect in the $P_{\rm II}-L_X$ relation has been argued by Treister et al. (2004) 
based on continuum of Great Observatories Origin Deep Survey,  but the highly 
spectroscopically complete deep and wide-area {\em Chandra} surveys support this relation 
(Barger et al. 2005, hereafter B05). This controversial issue motivates us to explore whether
there is a relation between the obscuring matter and AGN's engines.

The infrared emission  may provide new clues to look into the properties of 
the covering area. Infrared Space Observatory ({\em ISO}) data show that near- to mid-IR SEDs 
in PG quasars are likely due to reprocessing of dust heated by the central AGNs, even far IR 
band (Haas et al. 2000, 2003). 
Radio-quiet PG quasars usually have prominent big blue bumps (Elvis et al. 1994)
and hence they tend to have small inclinations to observers (Laor 1990). This is also supported
by the low $E_{\rm B-V}$ absorption and non-spherical distribution of dust (Haas et al. 2000).
The differences among these objects can be 
neglected in anisotropic properties of infrared (from compact tori) and optical ultraviolet 
emissions (from accretion disks). This allows us to reliably get the properties of torus 
obscuration and then to find evolutionary consequences of the tori from the 
multi-wavelength continuum.

In this Letter, we investigate  the covering factors of the tori from the multi-wavelength 
continuum. We show evidence for the agreement between the covering factors and the population 
fractions of AGNs. This implies that the AGN populations are changing with the evolution 
of the opening angle of the tori.

\section{The sample}
Following Granato \& Danese (1994), we define 
\begin{equation}
\crr=\frac{\int_{\rm IR}L_{\nu}d\nu}{\int_{\rm OUV}L_{\nu}d\nu},
\end{equation}
where $L_{\nu}$ is the continuum of the quasar, "IR" and "OUV" stand for the integral limits of the 
spectrum in infrared and optical-ultraviolet ($0.01-1\mu$m) bands, respectively. Since the uncertainties 
of extreme ultraviolet (EUV) spectrum, we get the EUV luminosity $L_{\rm EUV}(0.01-0.1\mu)\approx 2.2L_{0.1-1\mu}$
by simply extrapolating the spectrum of $F_{\nu}\propto \nu^{-0.5}$ (Rowan-Robinson 1995), where $L_{0.1-1\mu}$ 
is the luminosity integrated from $0.1$ to $1\mu$m and can be conveniently obtained from the observations. EUV 
luminosity has uncertainties less than 20\% from this extrapolation (Granato \& Danese 1994). We thus have 
$L_{\rm OUV}\approx 3.2L_{0.1-1\mu}$, 
with uncertainties less than $\sim 14\%$ due to EUV. Considering the debate of the IR origin from AGNs (e.g. 
Rowan-Robinson 1995; Haas et al. 2003), we use three bins as the main contribution from the torus: $L_{1-25\mu}$, 
$L_{1-60\mu}$ and $L_{1-100\mu}$, which are luminosities integrated for $1-25\mu$m, $1-60\mu$m and $1-100\mu$m, 
respectively, and have $\crr_{25}$, $\crr_{60}$ and $\crr_{100}$ subsequently.
We searched for radio-quiet PG quasars with $2-10$ keV X-ray and IR observations from the published literatures,
but we exclude some quasars based on the below criterion. 
The available data of infrared spectrum are from {\em IRAS} and {\em ISO} observations.
The sample is given in Table 1. The observed flux density is converted 
to monochromatic luminosity in the rest frame using $H_0=75~{\rm km~s^{-1}~Mpc^{-1}}$ and  $q_0=0.5$. 

\begin{table*}[t]
\begin{center}
\footnotesize
\centerline{\sc Table 1. The PG Quasar Sample (See the complete table in electronic version)}
\vglue 0.1cm
\begin{tabular}{lcccccccl}\hline\hline
Name  &$z$  &$\llnir$&$\llmir$&$\llfir$&$\llouv$&$\lvdip$&$\llx$&Ref.\\ 
(1)&(2) &(3)      &(4)  & (5)    &(6)     &(7)&(8) &(9) \\  \hline
PG0003$+$199 & $0.026$ & $44.51\pm0.03 $ & $ 44.57\pm0.04 $ & $< 44.59$ & $44.59\pm0.03 $& $14.50\pm0.09 $ & $43.06\pm0.02$ &11, 12, 4 \\
PG0026$+$129 & $0.142$ & $< 45.17 $      & $< 45.33 $       & $< 45.37$ & $45.37\pm0.04 $& $14.39\pm0.10 $ & $44.43\pm0.07$ &2, 12 ,12 \\ \hline
\end{tabular}
\end{center}
\end{table*}
\normalsize

We get the dip frequency via fitting the optical-ultraviolet spectrum 
as a power law and the near infrared spectrum (shorter than 5$\mu$m) as the Planck function
(Barvainis 1990, Kobayashi et al. 1993). When the parameter $\chi^2$ defined as
\begin{equation}
\chi^2=\sum_i^N\left[f_1\nu_i^{-\alpha}+f_2B\left(\nu_i,T_{\rm evap}\right)-f_{\nu_i}^0\right]^2,
\end{equation}
reaches a minimum value, we get the best-fitted spectrum 
$f_{\nu}=f_1\nu^{-\alpha}+f_2B\left(\nu,T_{\rm evap}\right)$ and hence the dip frequency of the spectrum 
around $1\mu$m. Here $T_{\rm evap}$ is the evaporation temperature of dust, 
$B\left(\nu_i,T_{\rm evap}\right)$ is the Planck function and $f_{\nu_i}^0$ is the flux at a frequency 
$\nu_i$. We take the evaporation temperature of dust $T_{\rm evap}=1500$K. 
The error bar of the dip frequency is then given by 
$\sigma_{\nu_{\rm dip}}^2=\left(\dd\nu_{\rm dip}/\dd f_1\right)^2\sigma_{f_1}^2
+\left(\dd\nu_{\rm dip}/\dd f_2\right)^2\sigma_{f_2}^2+\left(\dd\nu_{\rm dip}/\dd\alpha\right)^2\sigma_{\alpha}^2$, 
where $\sigma_{f_1}$, $\sigma_{f_2}$ and $\sigma_{\alpha}$ are the error bars 
of $f_1$, $f_2$ and $\alpha$ deduced from uncertainties of $f_{\nu_i}^0$.
The quality of the spectra of some objects around the dips is not good enough to obtain the dip frequency 
in the present sample. We exclude those objects with $\Delta \log \nu_{\rm dip}\ge 0.2$ and
get the exact frequencies of the dips in 34 quasars given in Tab 1. 

We should note that the near IR emission from host galaxies may have potential influence on 
the ratio of $\crr$ if the hosts are bright enough. A crude criterion for the host contribution is the ratio
of luminosities of quasar to host at $R-$ or $V-$band from the available data. It can be
obtained by extrapolating the OUV spectrum to $1\mu$ and getting $H-$band luminosity of the quasar from
its mean NIR spectrum of $f_{\nu}\propto \nu^{-1.3}$ (Haas et al. 2003) and $H-$band luminosity of host 
from $M_{\rm V}$ or $M_{\rm R}$ based on the spectrum of star light (Kriss 1988). 
We have the critical ratios of luminosities of quasars to hosts $L_{\rm V}^{\rm Q}/L_{\rm V}^{\rm H}\ge 3.4$ 
at $V-$band and $L_{\rm R}^{\rm Q}/L_{\rm R}^{\rm H}\ge 2.4$ at $R-$band, which correspond to
$L_{\rm H}^{\rm Q}\ge 2L_{\rm H}^{\rm H}$ at $H-$ band. Here the 
upper scripts Q and H refer to quasars and hosts. 
Though we exclude some quasars by the criteria, we have to keep in 
mind that the present sample could potentially cover other 
quasars (we are not able to distinguish them without available data of their host galaxies).
With these criteria, the uncertainties of the covering
factors are $\Delta {\cal C}_0/{\cal C}_0\le 0.05$ and $\Delta {\cal C}_1/{\cal C}_1\le 0.07$ and 
the uncertainties of the dip frequencies are $\Delta \log \nu_{\rm dip}\le 0.03$ due to the host's contaminations.
It should also be noted that the optical component extends past $1\mu$m to slightly contribute to IR flux,
which is neglected in this paper. We also note that the apertures are different for each bands (see
Table 1). Neugebauer et al. (1987) calibrated $1.3-10.1\mu$m photometry with respect to the standard stars. 
Haas et al. (2000; 2003) corrected the
source flux for the aperture size. We should use the total IR emission from the torus to determine the covering 
factor, however, the far IR emission and some of mid-IR in some AGNs could be powered by the extended torus, 
the narrow line region (e.g. NGC 1068; Bock et al. 2000), starburst or star from galactic stars. It is difficult 
to estimate the cutoff wavelength of the infrared emission from the torus, but three
ratios of $\crr_{25}$, $\crr_{60}$ and $\crr_{100}$ are used to show how the covering factor relies on the cutoff
wavelength. A crude estimation of the distance of dust to the center engine shows 
$R\le 4.7L_{45}^{1/2}\lambda_{25}^2$ pc from $aT^4\le L/4\pi R^2c$, where $L_{45}=L/10^{45}$erg s$^{-1}$,
$\lambda_{25}=\lambda/25\mu$, $a$ and $c$ are the black body radiation constant and the light speed, 
respectively, based on the assumption that the dust is heated by the central engine. Thus the ratios
of $\crr_{25}$, $\crr_{60}$ and $\crr_{100}$ correspond to the distances of dust, 
$4.7L_{45}^{1/2}\lambda_{25}^2$, $27.1L_{45}^{1/2}\lambda_{60}^2$ and 
$75.2L_{45}^{1/2}\lambda_{100}^2$ pc, respectively, where $\lambda_{60}=\lambda/60\mu$, 
$\lambda_{100}=\lambda/100\mu$. 

\section{Geometry of Torus and Statistics}

\subsection{Covering factors and AGN populations}
The geometry and physical conditions (e.g. dust and cloud properties) of the torus are 
not fully understood. Emergent spectra from the tori have been extensively calculated
for different geometry by Granato \& Danese (1994) and Efstathiou \& Rowan-Robinson (1995) stemming 
from the work of Pier \& Krolik (1992) and are able to generally fit the observed IR continuum. However, 
the IR optically thin grains can also produce the observed IR spectra of PG quasars (Barvainis 1990). 
This assumption is partially supported by the recent
observations of NGC 1068 of Jaffe et al. (2004), who find a spatially averaged peak silicate 
absorption depth, $\langle\tau_{\rm SiO}\rangle=0.3$ and $\langle\tau_{\rm SiO}\rangle=2.1$
in front of the $T=320$K and the hot components, respectively.
A clumpy model of torus has been proposed by Nenkova et al. (2002) and also gives reasonable
infrared SEDs. Though these uncertainties, the continuum 
may allow us to estimate the covering factors.

Assuming the isotropic emission from the central engine, Granato \& Danese (1994)
give the covering factor obtained from $\crr$ based on the energy balance 
\begin{equation}
{\cal C}=\frac{\crr}{1+\crr (1-p)}=\left\{
\begin{array}{ll}
\calc_0=\frac{\crr}{1+\crr}&(p=0),\\
                   &      \\
\calc_1=\crr               &(p=1),
\end{array}
\right.
\end{equation}
where $p$ is the ratio between the integrated flux emitted by the dust at the equator and that 
emitted at the pole. For $p=1$, the torus is essentially transparent to infrared radiation, whereas
the $p=0$ torus is optically thick to infrared photons. Though the emission from 
accretion disks is anisotropic (Netzer 1985), it is treated as a point of the illuminating
energy source in the popular models (Pier \& Krolik 1992; Granato \& Danese 1994; Efstathiou \& 
Rowan-Robinson 1995). The differences due to orientations could be neglected in the observed optical, 
ultraviolet and infrared luminosities among the objects in the present PG quasar sample.  Furthermore, 
it may be reasonable to assume that the torus and disk are co-plane, implying the orientation effects 
of disk and torus are similar. The parameter $\crr$ thus only weakly depends on the orientations. 

\begin{table*}[t]
\begin{center}
\footnotesize
\centerline{\sc Table 2. The Correlation Analysis}
\vglue 0.1cm
\begin{tabular}{lccccccccc}\hline\hline
            &\multicolumn{4}{c}{${\cal C}_0-L_X$}& &\multicolumn{4}{c}{${\cal C}_1-L_X$}\\ \cline{2-5}\cline{7-10}
            & $q$ & $k$ & $\rho$ & $p$              & & $q$ & $k$ & $\rho$ & $p$ \\ \hline
$\crr_{25}$ &$-0.36\pm0.07$ &$-0.17\pm0.04$ &$-0.53$ &$3.0\times 10^{-4}$ & &$-0.16\pm0.10$ &$-0.22\pm0.05$ &$-0.53$ &$3.0\times 10^{-4}$\\
$\crr_{25}^a$ &$-0.41\pm0.07$ &$-0.12\pm0.04$ &$-0.51^b$ &$1.7\times 10^{-3}$ & &$-0.22\pm0.09$ &$-0.16\pm0.05$&$-0.50^b$&$2.0\times 10^{-3}$\\
$\crr_{60}$ &$-0.42\pm0.08$ &$-0.10\pm0.04$ &$-0.33$ &$2.4\times 10^{-2}$ & &$-0.21\pm0.12$ &$-0.13\pm0.08$ &$-0.33$ &$2.4\times 10^{-2}$\\
$\crr_{100}$&$-0.32\pm0.09$ &$-0.13\pm0.05$ &$-0.36$ &$1.3\times 10^{-2}$ & &$-0.08\pm0.14$ &$-0.19\pm0.07$ &$-0.36$ &$1.3\times 10^{-2}$\\ \hline
\end{tabular}
\parbox{5.7in}
{\baselineskip 9pt
\noindent
{\sc Notes.} $\log {\cal C}=k\log L_{X,42}+q$, where $L_{X,42}=L_X/10^{42}$erg~s$^{-1}$. 
$\rho$ is the Spearman's coefficient and $p$ is the null-probability. $a$: the case excluding the upper limit sources.
$b$: refers to the Pearson's coefficient. There are too many upper limit sources in $\crr_{60}$ and $\crr_{100}$.
It should be noted the ASURV program using survival
statistics for the upper limits assumes that the limits are distributed
like the detections.
}
\end{center}
\end{table*}
\normalsize

We use the ASURV (Isobe et al. 1986)
to analyze the correlation for the censored data of the present sample. The results
are given in Table 2, which are shown in Fig 1{\em a} and 1{\em b}. We find that $\calc-L_X$ for 
${\crr_{25}}$ is much stronger than that for $\crr_{60}$ and $\crr_{100}$, especially $\calc_1-L_X$ for
$\crr_{25}$ nicely agrees with the fraction of type II AGNs 
(see below). The scatters in this plot may be caused by the complex geometry of the real torus 
and different values of $p$ in each quasars. We have checked whether the $\crr/{\cal C}-L_X$ correlation 
is introduced by that of $L_X$ and $L_{\rm OUV}$. Using the multivariate
correlation analysis, we find a strong correlation among $L_{1-25\mu}$, $L_X$ and $L_{\rm OUV}$
as $\log L_{1-25\mu}=6.53+0.98\log L_{\rm OUV}-0.14\log L_X$ with a Pearson's coefficient of $r=0.98$,
which can be translated into $\log \crr_{25}=6.53-0.02\log L_{\rm OUV}-0.14\log L_X$. This is consistent with 
the ${\cal C}-L_X$ for $\crr_{25}$ in Table 2 . Thus the  ${\cal C}-L_X$ relations here are true.
It is interesting to note the dependence of the correlation coeffients for ${\cal C}-L_X$ vs ${\cal R}$
for varoius IR cutoffs. This might imply a constarint on the extent to which the FIR is heated by AGN.

\vglue 0.5cm
\figurenum{1}
\centerline{\includegraphics[angle=-90,width=8.5cm]{f1.eps}}
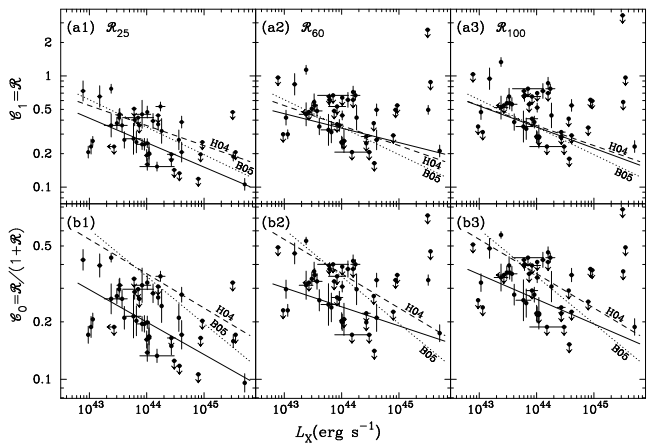
\figcaption{\footnotesize The plot of ${\cal C}-L_X$. The solid line
represents the best fit of the correlation by the least square method.
It is found that ${\cal C}_1-L_X$ relation for $\crr_{25}$ is consistent with
$P_{\rm II}-L_X$ relation found by {\em Chandra} and {\em HST} deep surveys.
The dashed and dotted lines are the fitted results from the surveys in Hasinger (2004) and Barger
et al. (2005), respectively. 
}
\label{fig1}
\vglue 0.5cm

With the geometry of the torus in Granato \& Danese (1994), we have
$\Delta \Omega/4\pi=\cos\theta =\calc$,
where $\Delta \Omega$ is the solid angle of the torus subtending the accretion disk
and $\theta$ is the half opening angle ($\le 90^{\circ}$). The deep surveys of {\em HST} and 
{\em Chandra} provide the fraction of type II AGNs, $P_{\rm II}=N_2/(N_1+N_2)$, where $N_1$ 
and $N_2$ are the numbers of type I and II AGNs, respectively. If the observer is 
located within the opening cone of the dusty torus, the object appears as type I, otherwise 
as type II. We thus have $P_{\rm II}=\Delta \Omega /4\pi$ allowing for the comparison with the 
results from surveys. The available data can be found in H04, who summarizes the 
results from {\em Chandra}, {\em HST} and {\em ASCA} (Ueda et al. 2003) based on the criterion 
of hydrogen column density. There are some type II AGNs missed in X-ray band, the 
$P_{\rm II}-L_X$ relation may be caused by the selection effect (Treister et al. 2004). 
In addition, some of AGNs in the deep {\em Chandra} field show some elusive properties in optical 
band (Maiolino et al. 2003), but a highly spectroscopic research on the
populations of {\em Chandra} deep field shows a robust $P_{\rm II}-L_X$ relation (B05).
We fit the data in Fig 6 in H04 and Fig 19 in B05 using the least square method and get,
\begin{equation}
\log P_{\rm II}=\left\{
\begin{array}{ll}
(-0.18\pm 0.02)+(-0.09\pm 0.01)\log L_{X, 42}& ({\rm H04}),\\
~                                               & ~          \\
(0.03\pm 0.50)+(-0.24\pm 0.34)\log L_{X, 42}& ({\rm B05}).
\end{array}
\right.
\end{equation}
It is very interesting to find the $P_{\rm II}-L_X$ relations in H04 and B05 nicely agree with 
each other within the uncertainties. This reflects some physical connections between the obscuring 
X-ray and the emission line regions in AGNs. Comparing eq. (4) with Table 2, we find that the present 
${\cal C}_1-L_X$ relation for $\crr_{25}$ is in agreement with the results of H04 and B05.

The agreement between the ${\cal C}_1-L_X$ for $\crr_{25}$ and the $P_{\rm II}-L_X$ relations indicates: 
1) a natural connection between the reprocessing matter (the compact torus) and the obscuring X-ray parts 
of AGNs (H04) as well as the 
optical emission line region (B05); 2) the dusty torus evolution results in the population changing 
of AGNs; 3) the ${\cal C}_1-L_X$ and $P_{\rm II}-L_X$ relations are intrinsic and thus greatly enhances 
the unification scheme of compact torus for AGNs, especially the ${\cal C}_1-L_X$ for $\crr_{60}$ and 
$\crr_{100}$ are much
scattered and significantly deviate from the relation H04 and B05. The present sample is different from 
that in H04 and B05, so the ${\cal C}_1-L_X$ for $\crr_{25}$ just 
strengthens the intrinsically physical connections of the torus evolution and population changing.  
It would be important if the $P_{\rm II}-L_X$ relation can be confirmed for the {\em same} sample in 
different ways. 

A self-regulation is set up between the sublimation of torus and the accretion process. A toy model
of the torus fueling the black hole gives $\dot{M}\propto {\cal C}^{-2}$, where $\dot{M}$ is the
accretion rate of the disk [see their eq. (9) and (15) in Krolik \& Begelman (1988)]. We thus have 
a simple relation of ${\cal C}\propto L_X^{-1/2}$ if $L_X\propto \dot{M}$. This theoretical prediction 
is much steeper than the observational results in this paper, but the trend of ${\cal C}-L_X$ relation
agrees with eq. (4) and ${\cal C}_1-L_X$ for $\crr_{25}$. A proper explanation of the ${\cal C}_1-L_X$ 
relation will depend on 
detail micro-physics of the torus and is beyond the scope of this Letter.
The evolution of the torus geometry could be controlled by the collisions among clouds inside the 
torus (Krolik \& Begelman 1988), which inevitably lead to detectable emissions in radio and sub-GeV bands 
(Wang 2004). In addition, the X-ray Baldwin effect of the iron narrow K$\alpha$ line can be derived 
from the changing population of type II AGNs based on the deep surveys, strongly implying  an 
evolutionary consequence of the tori (Zhou \& Wang 2005). 

\subsection{dip frequencies}
It is well known that the dip may be caused by dust evaporation and its frequency corresponds 
to the evaporation temperature (Sanders et al. 1989). The
observed $1\mu-$dip frequency may depend on both the orientation of the torus and the covering factor, 
but we focus on how it relies on $\calc$.  Fig 2{\em a} and 2{\em b} show this dependence. 
Using the ASURV, we obtain
\begin{equation}
\log \nu_{\rm dip}=\left\{
\begin{array}{l}
(14.55\pm 0.04)+(0.14\pm 0.07)\log \calc_0,\\
~      \\
(14.52\pm 0.03)+(0.11\pm 0.05)\log \calc_1.
\end{array}
\right.
\end{equation}
with Spearman's  $\rho=0.50$ and probability is $4.2\times 10^{-3}$, and $\rho=0.50$ and probability 
is $4.4\times 10^{-3}$, respectively.  The scatters in the plots may be caused mainly by  the quality 
of the data, inclination differences among these objects, the complicate structures of the torus and 
potentially the contamination of host galaxies at $1\mu$m.
However $\nu_{\rm dip}-\calc$ relation still shows that the dip frequencies are adjusted by covering 
factors of the torus. It could be qualitatively explained by the changes of the covering factors of 
the torus due to AGN evolutions. $\nu_{\rm dip}$ as a powerful probe of the covering factor could be
robust in future for good quality data and detail calculations of the radiation transfer in the
torus.

\figurenum{2}
\centerline{\includegraphics[angle=-90,width=8.5cm]{f2.eps}}
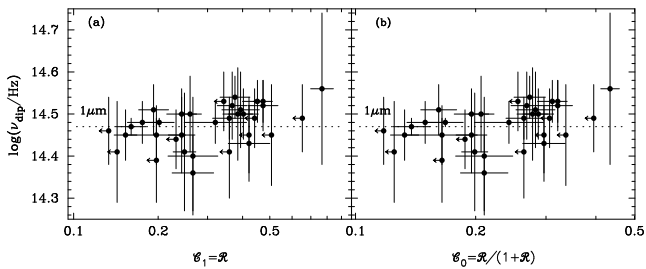
\figcaption{\footnotesize The covering factor $\calc$ and the dip frequency $\nu_{\rm dip}$.
This correlation shows an intrinsic connection between the torus and the accretion disk in AGNs,
implying a fueling process to a black hole.
}
\label{fig2}

\section{Conclusions and Discussions}
Fueling the black hole will change the covering factor of the torus.
It is thus expected that the covering factor could be a good indicator of the activity 
process of the black holes, showing the evolution of the active nucleus. 
Here we suggest an evolutionary sequence: higher $\cc$ $\rightarrow$ lower $\cc$. 
We find two clues to understanding the evolution: 1) changing populations of type II
AGNs as one consequence of AGN evolution due to increases of the opening angles; 2) the 
dip frequency increases with covering factor.  The two consequences could be used to 
trace the active processes of galactic nuclei statistically.

The $\nudip-\calc$ relation provides a new clue to understand the relation between the accretion 
disks and the tori. It reflects the inter region between the outer boundary of the disks and the 
inner edge of the torus. We poorly understand the physical processes in this region
as well as how to fuel the black hole on pc scale. 
Future observations of infrared interferometry telescopes could unveil something important taking 
place in this region, especially work on the survey sample would produce more robust results (Steffen 
et al. 2004). It is expected for {\em Spitzer} to make an attempt  for a sample of
deep surveys to construct more solid statistic properties of the tori.

\acknowledgements
The authors are grateful to the referee, R. Antonucci, for very helpful comments and constructive
criticism improving the manuscript. S. N. Zhang is thanked for discussions. This research is supported 
by a Grant for Distinguished Young Scientists from NSFC-10325313, NSFC-10233030 and 973 project.

\clearpage

\clearpage

\clearpage

\clearpage


\begin{thebibliography}{}
\bibitem[]{417}Antonucci, R. R. J., 1993, ARA\&A, 31, 473
\bibitem[]{418}Antonucci, R. R. J. \& Miller, J. S., 1985, ApJ, 297, 621
\bibitem[]{419}Barger, A. J., et al., 2005, AJ, 129, 578
\bibitem[]{421}Barvainis, R., 1990, ApJ, 353, 419
\bibitem[]{422}Bock, J. J., et al., 2000, AJ, 120, 2904
\bibitem[]{423}Cao, X., 2005, ApJ, 619, 86
\bibitem[]{424}Efstathious, A. \& Rowan-Robinson, M., 1995, MNRAS, 273, 649
\bibitem[]{425}Elvis, M., et al. 1994, ApJS, 95, 1
\bibitem[]{426}George, I. M., et al. 2000, ApJ, 531, 52
\bibitem[]{427}Granato, G. L. \& Danese, L., 1994, MNRAS, 268, 235
\bibitem[]{428}Haas, M., et al. 2000, A\&A, 354, 453
\bibitem[]{429}Haas, M., et al. 2003, A\&A, 402, 87
\bibitem[]{430}Hasinger, G., 2004, Nucl. Phys. B (Proc Suppl.), 132, 86
\bibitem[]{431}Isobe, T. Feigelson, E. D. \& Nelson, P. L., 1986, ApJ, 306, 490
\bibitem[]{432}Jaffe, W., et al., 2004, Nature, 429, 471
\bibitem[]{434}Kobayashi, Y., et al., 1993, ApJ, 404, 94
\bibitem[]{435}Kriss, G.A., 1988, ApJ, 324, 809
\bibitem[]{436}Krolik, J.H. \& Begelman, M.C., 1988, ApJ, 329, 702
\bibitem[]{437}Laor, A., 1990, MNRAS, 246, 369
\bibitem[]{440}Lawson, A. J., \& Turner, M. J. L., 1997, MNRAS, 288, 92
\bibitem[]{442}Maiolino, R., et al. 2003, MNRAS, 344, L59
\bibitem[]{445}Nenkova, M., Ivezic, Z. \& Elitzur, M., 2002, ApJ, 570, L9
\bibitem[]{446}Netzer, H., 1985, MNRAS, 216, 63 
\bibitem[]{448}Neugebauer, G., et al., 1987, ApJS, 63, 615
\bibitem[]{449}Osterbrock, D.E. \& Shaw, R.A., 1988, ApJ, 327, 89
\bibitem[]{450}Page, K. L., O'Brien, P. T. et al. 2004, MNRAS, 347, 316
\bibitem[]{451}Piconcelli, E., et al., 2004, astro-ph/0411051
\bibitem[]{452}Pier, E. A. \& Krolik, J. H., 1992, ApJ, 401, 99
\bibitem[]{453}Porquet,D., et al. 2004, A\&A, 422, 85
\bibitem[]{454}Reeves, J. N. \& Turner M. J. L. 2000, MNRAS, 316, 234
\bibitem[]{455}Sanders, D.B., et al., 1989, ApJ, 347, 29
\bibitem[]{456}Steffen, A.T. et al., 2003, ApJ, 596, L23
\bibitem[]{457}Steffen, A.T. et al., 2004, AJ, 128, 1483
\bibitem[]{459}Tovmassian, H. M., 2001, Astron. Nachr. 2, 87 
\bibitem[]{460}Treister, E., et al., 2004, ApJ, 616, 123
\bibitem[]{462}Ueda, Y., et al., 2003, ApJ, 598, 886
\bibitem[]{463}Wang, J.-M., 2004, ApJ, 614, L21
\bibitem[]{464}Wilkes, B. J., et al. 1999, ApJ, 513, 76
\bibitem[]{465}Zhang, S. N., 2005, ApJ, 618, L79
\bibitem[]{466}Zhou, X.-L. \& Wang, J.-M., 2005, ApJ, 618, L83
\end{thebibliography}
\end{document}